# NEW PARADIGMS IN MAGNETIC RECORDING

BY MARTIN L. PLUMER, JAN VAN EK, AND WILLIAM C. CAIN

The phenomenal increase in the storage capacity of magnetic hard disc drives (HDD) in recent decades has been fueled not only by clever improvements in the engineering of tiny devices but also by discovery, and advances in understanding, of fundamental physical phenomena associated with magnetism at the nanometer length scales [1,2]. Over the past 55 years, the Areal Density (AD) has increased from kb/in², to Mb/in², to Gb/in², and recently to Tb/in². The nontrivial task of making the early inductive transducers smaller (aided by the advent of thin-film technology) had enabled Compound Annual Growth Rates (CAGR) of about 40% in the first 35 years. Subsequent increases in AD were largely due to fundamental changes in the three main magnetic components of the hard drive: The recording media, the write element of the transducer and the read element of the transducer [3]. These three components, which are the focus of this review, interact and govern an important characteristic of information storage and retrieval in the HDD, namely the magnetic signal-to-noise ratio (SNR). Improvements in non-magnetic component technologies, such as signal processing, spindle bearings, active transducer-to-media spacing control, servo mechanical, and variable bit aspect ratio along the radial direction have also been crucial to AD growth.

The award of the Nobel Prize in Physics to Albert Fert and Peter Grünberg in 2007 for the discovery of Giant Magnetoresistance (GMR) indicates the importance of fundamental research to the HDD industry. GMR was incorporated into transducers for magnetic recording in the late 1990s and followed the use of ordinary Anisotropic Magnetoresistance (AMR) introduced earlier in that decade. The implementation of Tunneling

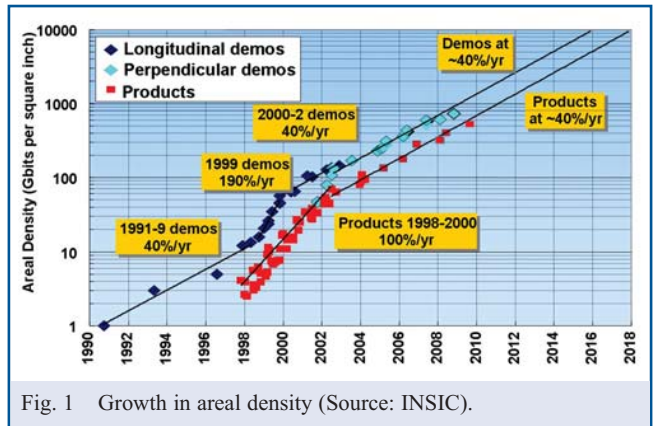

Fig. 1    Growth in areal density (Source: INSIC).

Magnetoresistance (TuMR) in 2005-6 has proven even more effective in increasing read-back amplitude [4].

The growth in CAGR from 40% to 60% to 100% which began in the mid 1990s and spanned the following several years (Fig. 1) was only partly due to these advances in the transducer read element. Significant improvements in the write element, and even more importantly in the recording media, were essential in facilitating this unprecedented shrinkage in bit size. Typically, the limiting factor in magnetic SNR is the noise from the media component.

The introduction of new technology into the recording system has occasionally resulted in large, although temporary, gains in CAGR. Over the past five years innovations such as the use of perpendicular recording have allowed for continued growth in AD although at more moderate and historic rates of 40-50%. Maintaining AD growth will continue to rely on the discovery and successful implementation of new concepts to improve magnetic sensors and storage media.

Fig. 1 also reveals that there has historically been a large gap of two years (and up to a factor two or three in AD magnitude) between laboratory demonstrations ('demos') and shipped product..

The recording layer in modern HDD discs is a granular Cobalt-Chrome-Platinum-based alloy with high uniaxial magnetocrystalline anisotropy derived from its hcp crystalline structure. Thin films 15 – 20 nm thick are composed of crystalline grains which are physically isolated through the segregation of inter-granular non-magnetic

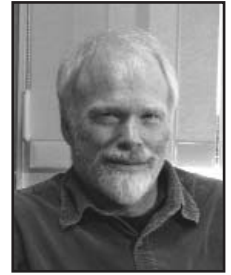

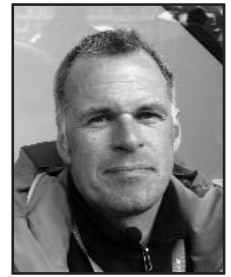

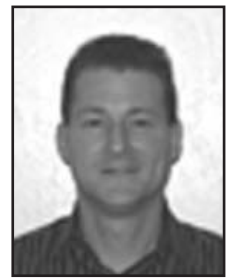

M.L. Plumer
<plumer@mun.ca>,
Memorial University
of Newfoundland,
St. John's, NL
A1B 3X7

J. van Ek, Western
Digital Corporation,
5863 Rue Ferrari,
San Jose, CA

W.C. Cain, Western
Digital Corporation,
3353-3355 Michelson
Dr., Irvine, CA

## SUMMARY

The magnetic hard disc drive industry continues to face serious challenges in its quest for ever decreasing bit size. This review summarizes recent advances and promising new technology which have foundations in fundamental physical principles. Some advantages of these new ideas are illustrated through micromagnetic modeling and the numerous challenges associated with their implementation are highlighted.

Cr-rich [5] and oxide materials [6] to the grain boundaries. Prior to 2006-7, all HDDs used so-called longitudinal media where the Co c-axis lies in the media plane. Prior to 2006-7, all HDDs used so-called longitudinal media where the Co c-axis lies in the media plane. The easy-axis anisotropy direction and magnetization were also in the film plane. Smaller grain size is a very important factor in increasing media SNR from a purely statistical point of view since SNR $\sim \sqrt{N}$ where N is the number of grains per bit. In the past decade, modal grain sizes have decreased from about 20 nm to about 9 nm [7].

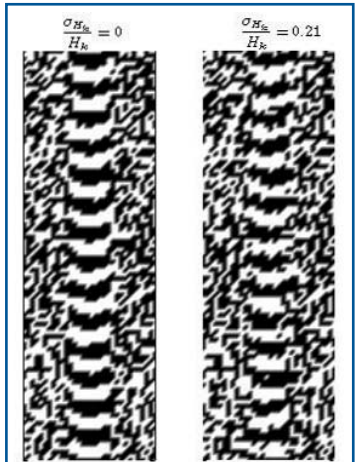

In addition to smaller grains, smaller grain-size distributions, as well as smaller variations in magnetic properties, are also essential for good media SNR. Media with crystalline anisotropy fields, $H_K$, which vary widely from grain-to-grain (for example) provide poor SNR as seen in the micromagnetic modeling results of Fig. 2. Correspondingly detrimental results are found if the grain size increases.

Fig. 2   Micromagnetic simulation results of recorded transitions at 788.8 kfci (kilo flux-changes-per-inch) showing detrimental effects due to an anisotropy distribution.

## SUPERPARAMAGNETISM AND THE TRILEMMA

Although the phenomenon of superparamagnetism was well known within the HDD industry for many decades, it was not until the early and mid-2000s that 'spontaneous' media grain magnetization reversal due to thermal fluctuations became a pressing issue. The large uniaxial anisotropy constant, $K_u = \frac{1}{2}H_K M$, of Co-based media inhibits the reversal of the grain magnetization vectors within a written bit. The energy barrier between +/- directions of $M$ is proportional to $E_B = K_u V$ where V is the grain volume $\sim d^3$ with d being a typical grain diameter [3]. Grain magnetization flipping is a thermal activation process governed by the Neel-Arrhenius law giving a time constant

$$\tau^{-1} = f_0 \exp(-E_B / k_B T)$$

where $f_0$ is the attempt frequency determined by intrinsic magnetic properties and is of the order $10^9 - 10^{12}$ Hz.

The superparamagnetic trilemma involves grain size, media anisotropy and the write-element magnetic field. In order to assure stored information does not degrade through spontaneous magnetization reversal, magnetic media with sufficiently large $E_B$ is required. In order to maintain adequate media SNR with smaller bits, smaller grains are necessary. This implies that there must be a concomitant increase in media anisotropy to prevent superparamagentic data loss. Thus, in order to purposely reverse grain magnetizations during the write process, larger head fields are required, typically about $H_K$. In order to ensure the stability of recorded transitions it is estimated that the parameter $K_u V / k_B T$ should be greater than about 60 at operating HDD temperatures of about 340 K. This then defines a relationship between grain size and the media anisotropy, as illustrated in Fig. 3.

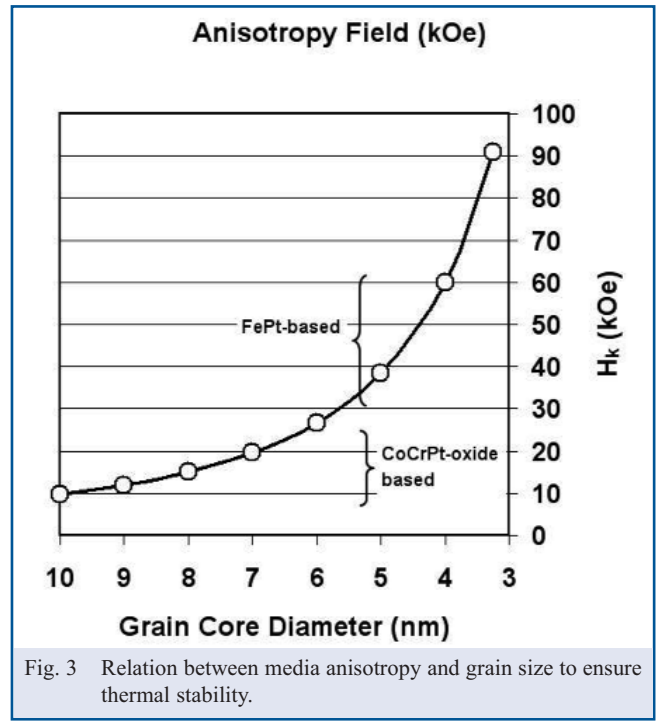

Fig. 3   Relation between media anisotropy and grain size to ensure thermal stability.

Large increases in write-element head fields were enabled throughout the 1990's through the use of higher magnetic moment materials placed at 'the business end' of the transducer. Ni/Fe Permalloy with a saturation magnetization of 1.0 T was replaced by Co/Ni/Fe alloys with 1.3-2.1 T and finally by CoFe having 2.4 T in the late 1990's. Unfortunately, the increases stopped there. No higher moment materials which are amenable for use in the write element are available and only modest enhancements in the 'head field' can be achieved by design tweaks.

Other options involving fundamental design changes have been implemented in recent years. These and a number of the more promising proposals for future HDD technology are reviewed in the following.

## PERPENDICULAR RECORDING

By changing sputtering conditions and substrates, the crystallographic c-axis, and hence easy magnetic axis, of Co-based thin films can be changed from in-plane to out-of-plane. Although known since the early days of HDD development, the longitudinal mode prevailed until 2005 [3,8-11]. There are five principal benefits which result from aligning bit magnetic moments perpendicular to the film plane compared to the longitudinal configuration.

1) The biggest advantage comes from the fact that the write field vector should be mainly directed perpendicular to the media plane. In perpendicular recording, the media is effectively in the write gap so that much larger fields are possible. This is achieved by adding a magnetically soft (low anisotropy) layer below the hard recording layer in the media which effectively then becomes part of the write element.

2) In addition to this substantial benefit, the magnetostatic field arising from neighboring bits in longitudinal recording tends to destabilize the transition (lower the energy barrier). The opposite is true in perpendicular recording. This can be easily understood by playing with a pair of bar magnets.

3) Whereas for longitudinal media the easy axis distribution is essentially 2D-isotropic, the easy axes for perpendicular media are contained within a cone of about 5 degrees. Smaller anisotropy distributions lead to better media SNR.

4) Grains in the perpendicular media are separated with oxide material resulting in sharper transition between ferromagnet/non-magnetic material (oxide)/ferromagnet materials. This will result in an improved switching field distribution of grains in perpendicular media. In longitudinal media Cr segregation is achieved by heating media ~250C so Cr composition profile varies over much wider distance. On the other hand, perpendicular media is fabricated at room temperature where metal oxide mixing can be avoided.

5) A final benefit from perpendicular recording is due to the fact that the stray field emanating from transitions is larger than in longitudinal recording. This yields a larger reader response and larger change in voltage from the read element giving a larger electronic contribution to overall SNR.

In the absence of the transition to this new technology, AD growth rates would probably have fallen well below the historical 40-50% enjoyed in recent years (Fig. 1).

## TUNNELING MAGNETORESISTANCE

In contrast with AMR, where intrinsic magnetic properties of a material result in a resistance change dependent upon the relative orientation of the magnetic moment and direction of electric current, both GMR and TuMR involve spin-dependent electron transport of two ferromagnetic layers separated by a thin non-magnetic film, as shown in Fig. 4. One of the ferromagnetic layers, the pinned layer (PL) has its moment's direction pinned through coupling to an (Ir-Mn or Pt-Mn) antiferromagnet, and the other free layer (FL) rotates in response media transition fields. Device stability is enhanced by the placement of Co-Pt-based permanent magnets adjacent the spin valve.

In GMR spin valves, the non-magnetic spacer is highly conductive and usually made of Cu. The current flows in the plane of the device and gives rise to spin-dependent electron scattering involving both bulk and surface electron states [3]. Typically GMR devices yield MR ratios of 10-15 %.

In contrast, the TuMR effect (first reported in 1975 [12]) involves current flow perpendicular to the magnetic films across an insulator (barrier). The effect is due to spin-dependent electron tunneling. Here, the difference in resistance from low (FL and PL moments parallel) to high (FL and PL moments antiparallel) resistance states is due to a difference in the spin-dependent density of states at the Fermi level of the two layers [3,4].

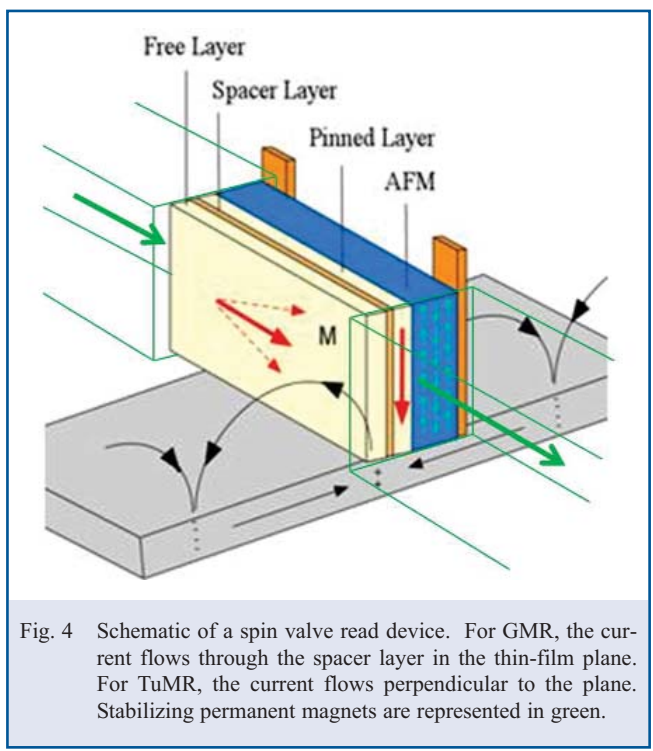

Fig. 4    Schematic of a spin valve read device. For GMR, the current flows through the spacer layer in the thin-film plane. For TuMR, the current flows perpendicular to the plane. Stabilizing permanent magnets are represented in green.

Today TuMR-based devices which have been incorporated into HDDs exhibit MR ratios of approximately 100% [4]. The barrier material of choice is MgO. Even higher MR ratios can be achieved at low temperatures and through ultra-high-vacuum epitaxial deposition techniques but these are not yet practical for low-cost, high-throughput manufacturing requirements. The production of sputter deposited ultra-thin films with reproducible electrical properties is challenging. This issue is exacerbated by the fact that tunneling properties depended strongly on the materials composition and roughness at surfaces, both of which are difficult to control in sputter deposition processes.

Unfortunately, the large increases in TuMR amplitudes relative to GMR have not resulted in the same improvements in electronic SNR. In addition to (limited) contributions from 1/f noise and Shot/Johnson noise, the dominant noise source is magnetic noise. Magnetic noise is caused by imperfect pinning of the pinned layer and by thermal fluctuations in the atomic spins of all the magnetic layers. These noise sources place demanding limits on the resistance-area (RA) product of the tunnel junction to be less than 1-2 $\Omega$-$\mu$m$^2$ [13,14].

## ENERGY ASSISTED MAGNETIC RECORDING

While perpendicular magnetic recording has been firmly established as the vehicle for mainstream mass produced disc drives, the trilemma is looming again. The requirement for smaller grain size in the media for good SNR, yet adequate thermal stability, implies increasing media anisotropy to the point where the recording process becomes marginal.

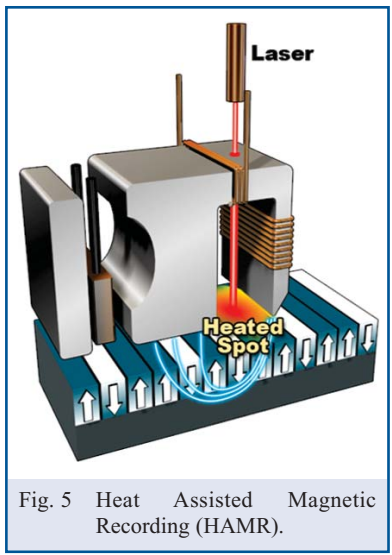

**Fig. 5** Heat Assisted Magnetic Recording (HAMR).

There are a number of ambitious proposals being explored to enable continued AD growth. Two of them can be classified under Energy Assisted Magnetic Recording (EAMR). In addition to a magnetic field from the write pole, another source of energy assists in the recording process. The most promising is Heat Assisted Magnetic Recording (HAMR) for application beyond 1Tb/in$^2$ [15], as illustrated in Fig. 5.

Due to the very small grain sizes, future medium anisotropy field H$_K$ is expected to be on the order of 50 kOe. With HAMR, the reversal of the grain magnetization is facilitated by raising the temperature of the media to about its Curie temperature. As the media cools the write field is applied to freeze in the magnetization in the desired orientation. Delivery of the heat will take place through a laser system whereby the laser spot is focused down to just below the track width (< 50 nm). The final shape of the light spot is obtained though the use of a near field transducer (where the laser light excites surface plasmons). At the end of the transducer an evanescent wave is produced which couples into the media and creates heat. Micromagnetic simulation results demonstrate significant SNR gains [16].

Also under consideration is Microwave Assisted Magnetic Recording (MAMR) where a microwave frequency 'assist' field is added to the static write field with some preferred orientation and polarization relative to the normal static write field [17]. The magnetization of a grain can then be reversed with a static write field that normally would be too small to cause reversal. Initial micromagnetic simulations show promising results albeit for relatively large magnitude of the oscillating field.

## BIT PATTERNED MAGNETIC RECORDING

A conceptually straightforward solution for the thermal stability problem at high areal density is to define the individual bits as continuous single domain entities with a relatively large volume and therefore a low anisotropy requirement. The bits or 'dots' on bit patterned media (BPM) [18] would be separated by non magnetic grooves, as illustrated in Fig. 6. This should be contrasted with the conventional approach where bits composed of many tiny grains. For instance, at 1Tb/in$^2$ the dots could be on the order of 13 × 13 nm$^2$ and about 10 nm thick. Assuming a single magnetic domain structure, this would provide thermal stability at very modest anisotropy values. However, the very low bit width-to-length ratio will have severe implications for the requirements on the servo tracking system and may result in undesirably low data throughput rates.

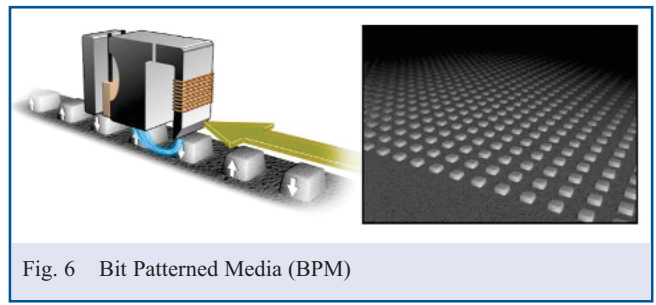

**Fig. 6** Bit Patterned Media (BPM)

With velocities of the write element relative to the media in the range 15 to 30 nm/ns, a 12 nm separation of the islands implies that the magnetic write field needs to be fully reversed in 0.8 to 1.7 ns. In itself this seems attainable, however small random variations in rotational speed of the disc, placement of the dots during disc manufacturing, and variability in magnetic properties of the patterned bits will introduce errors in the recording process, *i.e.* bits will be 'missed'. Correcting for this would require a very sophisticated high speed closed-loop feed back system that samples magnetic and spatial properties of upcoming islands and couples that together with information about disc spindle speed for compensation of the recording timing.

## OUTLOOK

In order to achieve the approximate 40% compound areal density growth rate that the HDD industry has delivered over the past 50 years, several key technology innovations have been employed. Many of the innovations in the last decade have been aided by fundamental materials science breakthroughs in head and media technology such as GMR and TuMR read head materials and AFC coupled longitudinal and granular oxide perpendicular media.

The most recent enabling technology, perpendicular magnetic recording, has allowed a rapid increase in areal density from 130 to 520 Gb/in² in under four years. As perpendicular recording technology moves up the "S" curve of maturity, the industry is focused on the next set of innovations that will continue to spark future areal density growth.

The continuing technical challenge in increasing HDD areal density is to achieve a balance among the signal-to-noise ratio and thermal stability of small grain media and the ability of the head to write the media while continuing to maintain the low cost per GByte needed to support unit growth.

Understanding and exploration of the large variety of fundamental physical phenomena involved in current, proposed and future technologies can only serve to enhance the possibilities for finding solutions to these challenges.

## ACKNOWLEDGEMENTS

We thank M. Steinitz for suggesting this review and H. Richter, D. Weller, A. Torabi, E. Champion and the referee for helpful comments and suggestions.